\documentclass[aps,pre,11pt,showpacs,superscriptaddress,tightenlines]{revtex4}

\usepackage{amssymb,amsmath,graphicx}

\begin{document}

\title{Desynchronization of systems of coupled Hindmarsh-Rose oscillators}

\author{A. Gjurchinovski}

\email{agjurcin@pmf.ukim.mk}

\affiliation{Institute of Physics, Faculty of Natural Sciences and Mathematics, Sts.\ Cyril and Methodius University, P.\ O.\ Box 162, 1000 Skopje, Macedonia}

\author{V. Urumov}

\email{urumov@pmf.ukim.mk}

\affiliation{Institute of Physics, Faculty of Natural Sciences and Mathematics, Sts.\ Cyril and Methodius University, P.\ O.\ Box 162, 1000 Skopje, Macedonia}

\author{Z. Vasilkoski}

\email{Zlatko_Vasilkoski@hms.harvard.edu}

\affiliation{Harvard Medical School, Goldenson 343, Boston, MA 02115, USA}


\begin{abstract}
It is widely assumed that neural activity related to synchronous rhythms of large portions of neurons in specific locations of the brain is responsible for the pathology manifested in patients' uncontrolled tremor and other similar diseases. To model such systems Hindmarsh-Rose (HR) oscillators are considered as appropriate as they mimic the qualitative behaviour of neuronal firing. Here we consider a large number of identical HR-oscillators interacting through the mean field created by the corresponding components of all oscillators. Introducing additional coupling by feedback of Pyragas type, proportional to the difference between the current value of the mean-field and its value some time in the past, Rosenblum and Pikovsky (Phys. Rev. E \textbf{70}, 041904, 2004) demonstrated that the desirable desynchronization could be achieved with appropriate set of parameters for the system. Following our experience with stabilization of unstable steady states in dynamical systems, we show that by introducing a variable delay, desynchronization is obtainable for much wider range of parameters and that at the same time it becomes more pronounced.  
\end{abstract}

\date{30 May 2011}

\maketitle

\section{Introduction} 

Multidisciplinary approach through modelling, computation and engineering has brought significant progress in many fields. Such is the case with the modern developments related to the Parkinson disease. The pathology is most easily recognized by the involuntary tremor of the patients limbs. Other key characteristics are rigidity of the posture, slow movements and postural instability. Condensed description of the topic can be found in the exposition by Schiff \cite{SCH10}. It has been observed, although not in all cases, that the neurons in the affected area of the brain responsible for movement control are to a significant extent synchronized in their bursting activity. The chalenge is to find ways to diminish the degree of synchrony. There are three main approaches: pharmacological therapy, introduction of surgical lesions in specific locations of the brain and deep brain stimulation (DBS). The latter technique was introduced into medical practice more than a decade ago. It consists of implantation of device providing electromagnetic perturbation with appropriate frequency to the affected brain area, but the physiological mechanism behind such clinical practice is not yet clearly understood.  

Synchronization in general, and in particular, synchronization in systems of large number of coupled oscillators is a topical area of research in the context of nonlinear sciences 
\cite{SinhrBOOK}. Although the synchronization phenomenon could play a constructive role
and its existence could cause a variety of useful applications, in some circumstances, as already mentioned, it is desirable to control the degree of  synchronization of the oscillator system from the outside, and in the limiting case, to desynchronize the system and to sustain such a state 
for a certain time interval. Synchronization of bursting neurons with delayed synapses has been discussed by Buri\'c et al. \cite{BUR08}. Computational studies were performed to find optimal waveforms for DBS \cite{FENG07}, to determine the influence of applying the input in different target zones \cite{PIR09}, several linear and nonlinear methods to achieve desynchronization were proposed by Tass et al. \cite{TASS08}. In this paper, we discuss the possibility of desynchronization in a population of globally coupled Hindmarsh-Rose oscillators \cite{HR84} by applying a variable-delay feedback control.

\section{System of interacting Hindmarsh-Rose oscillators}

An effective method to suppress the synchrony in a network of
globally coupled oscillators was proposed in 2004 by
Rosenblum and Pikovsky \cite{RP04}. They applied 
the time-delayed feedback control by taking the control signal 
to be the difference between the current value of the mean field, 
and its past (time-delayed) value (delayed mean field).
The analysis for different models of coupled bursting neurons 
showed that such a synchronization control is quite effective,
and it could be used to control (suppress) the pathological rhythms
in an ensemble of coupled neurons by a proper choice of the
feedback control parameters.
The main advantage of the proposed control method is the fact that
it requires neither information on the details of the individual 
oscillators and their interactions, nor access to their parameters. 
The method is also noninvasive from a control theory viewpoint,
since the feedback signal vanishes after the suppression
is achieved, although the technique is invasive in medical sense 
since it requires implantation of batteries and electrodes in the body of the pacient.

In our analysis, we have chosen the Hindmarsh-Rose system of
oscillators, which can be considered as a physiologically realistic model for describing the neuronal activity of the brain cells. In this collective model it is assumed that the oscillators 
are globally coupled through their mean field.

We consider a system of $N$ identical Hindmarsh-Rose oscillators \cite{HR84}, in which the dynamics of the individual neurons is described by the following equations:
\begin{eqnarray}
\dot x_i &=& y_i-x_i^3+3x_i^2-z_i+3+F_1(t)+F_2(t),\\
\dot y_i &=& 1-5x_i^2-y_i,\\
\dot z_i &=& 0.006\left[4(x_i+1.56)-z_i\right].
\end{eqnarray}
Here $x_i$ is the membrane potential, $y_i$ and $z_i$ represent the fast and slow currents of ions for the $i$-th oscillator, while the constant term in the first equation is the external current. 
The coefficients in equations (1)--(3) are commonly selected values for the parameters modeling the slow and fast ion channels.
The term $F_1(t)$ describes the global coupling between the oscillators,
and is given by:
\begin{equation}
F_1(t)=K_{MF} X(t),
\end{equation}
where
\begin{equation}
X(t)=\frac{1}{N}\sum_{i=1}^{N} x_i(t).
\end{equation}
is the mean field, created by the $x$-components of all oscillators 
($i=1,2,\dots,N)$. The term $F_2(t)$ is the variable-delay feedback force defined by the 
difference between the delayed and the current mean field signal:
\begin{equation}
F_2(t)=K\left[X(t-\tau(t))-X(t)\right],
\end{equation}
Here $\tau(t)$ represents the delay which is taken to be variable and dependent on the time $t$. The structure of the feedback term $F_2(t)$ is identical to the form first proposed by Pyragas 
\cite{PYR92} for stabilization of unstable periodic orbits in chaotic systems. 
In the absence of control $(F_2=0)$, the synchronous state onsets when the coupling strength 
$K_{MF}$ exceeds a certain critical value. The increase of the coupling parameter $K_{MF}$ beyond the 
critical value manifests itself via the appearance of macroscopic oscillations of the mean field $X(t)$, which models the pathological brain activity. The goal is to achieve desynchronization of the oscillators population, i.e. to suppress the mean field $X(t)$ by applying the feedback force
$F_2(t)$ and making a suitable choice of the control parameters.

The motivation to use variable time-delay comes from previous studies of ours \cite{GU08} with varying delay applied to problems of stabilization of unstable steady states in various systems \cite{HS05,DHS07,YWH06}. The results were very favourable leading to significant enlargement of the domains in the parameter space for which stabilization is achievable. Moreover the approach to the stationary state is faster and its stability becomes more robust.  

\begin{figure}
\includegraphics[width=0.8\columnwidth,height=!]{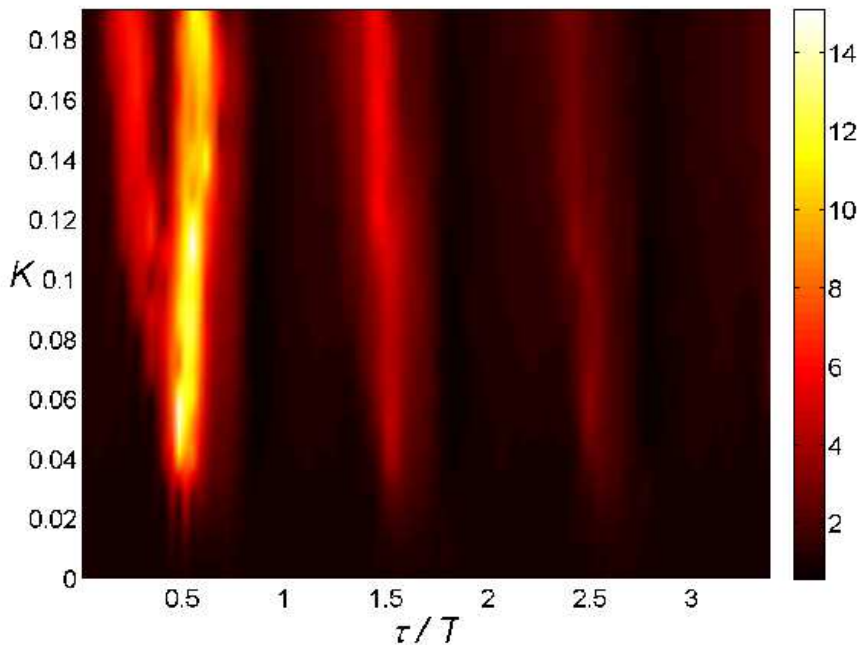}
\caption{\label{sl38}
Numerical calculations of the suppression coefficient 
$S$ of the mean field oscillations in a system of 1000 
globally coupled Hindmarsh-Rose oscillators in the plane
parametrized by the feedback strength $K$ and the time delay
$\tau$. The delay is normalized by the average period 
$T=175$ of the mean-field oscillations in the system without control. 
The value of the coupling strength is $K_{MF}=0.08$. 
The time delay $\tau$ in the feedback force is constant (Pyragas method). 
Values of $S$ are given in a grayscale coding. 
The domain of suppression consists of islands located around
$\tau/T\approx(2n+1)/2$, $n=0,1,\dots$. 
For $\tau=0$ control can not be achieved, meaning that desynchronization 
is not possible without a feedback controller.}
\end{figure}

To characterize the influence of the feedback gain parameter $K$ and the time delay $\tau$ in controlling collective synchrony of the Hindmarsh-Rose oscillators, we use the suppression coefficient defined as:
\begin{equation}
S=\sqrt{\frac{\mathrm{var} (X)}{\mathrm{var} (X_f)}},
\end{equation}
where $X$ and $X_f$ are the mean fields in
the absence and presence of the feedback, respectively,
and 
$\mathrm{var} (X)$ 
(analogous for $X_f$) is the variance of the mean field $X$. 
Results of numerical calculations for the dependence of the suppression factor
$S$ on the control parameters $K$ and $\tau$ in the case when the delay is constant are shown in Fig. \ref{sl38}. The simulation is performed for $N=1000$ oscillators for the strength of the 
internal coupling $K_{MF}=0.08$. The values of the suppression factor $S$ are given in a grayscale 
coding. The values of the delay $\tau$ in the horizontal axis are 
normalized by the average period $T$ of the mean-field oscillations
without control ($T\approx175$). 
The domain of suppression consists of several islands encompassing
the values of $\tau/T$ equal to $(2n+1)/2$, where
$n=0,1,\dots$ 
The maximum value of the suppression coefficient in the depicted
range of the control parameters is about 15. 

\begin{figure}
\includegraphics[width=0.8\columnwidth,height=!]{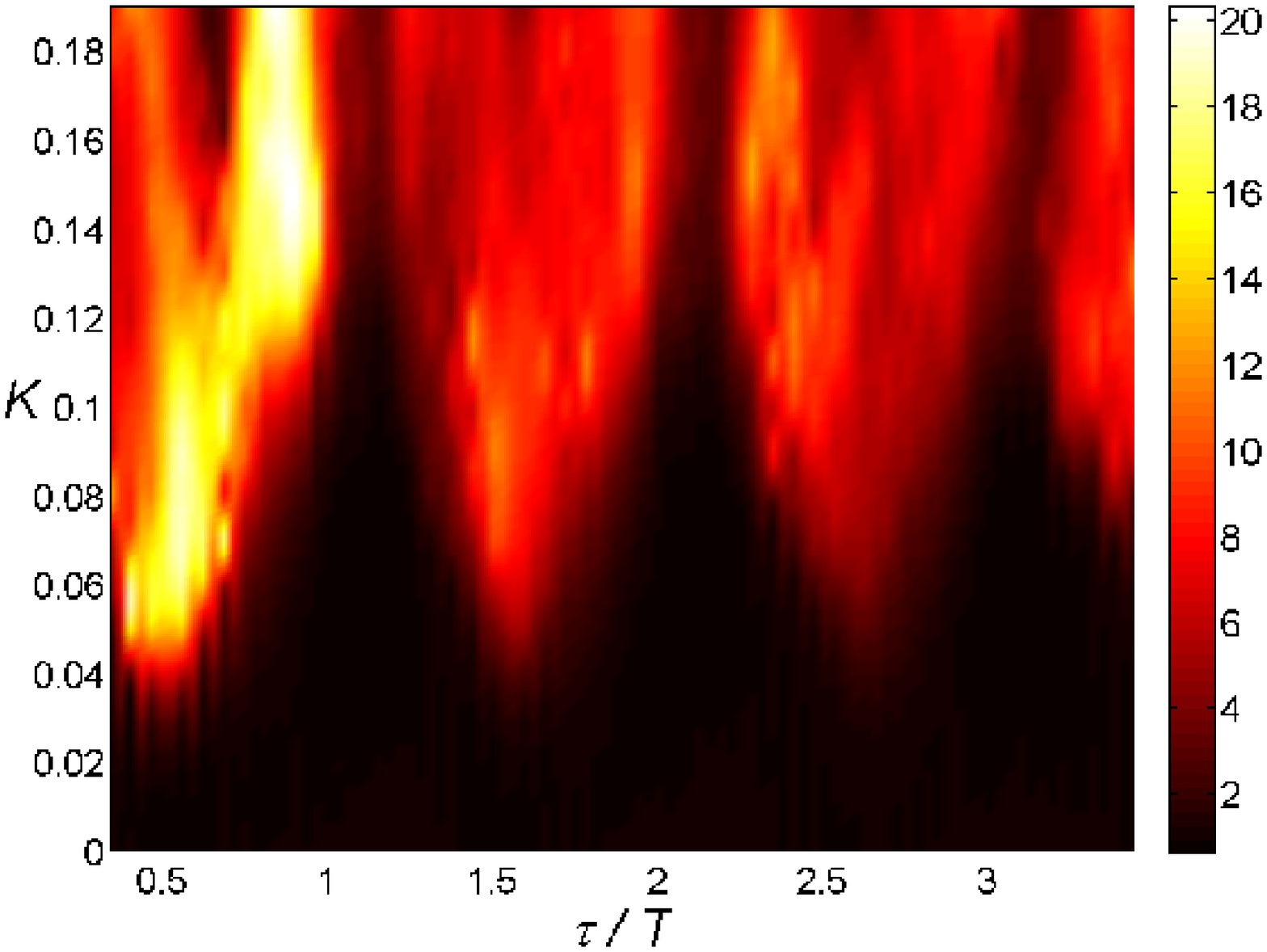}
\caption{\label{sl39}
The extension of the domain of suppression by a variable-delay feedback control.
The modulation of the delay is with a sine wave with amplitude
$\varepsilon=40$ and frequency $\nu=10$.
The rest of the parameters are as in Fig. \ref{sl38}. Notice the shift of the origin due to the restriction $\tau\ge\varepsilon$.}
\end{figure}

To investigate numerically the effects of inclusion of a variable delay
in the above scheme for desynchronization of globally coupled
oscillators, we will use a deterministic delay modulation in a form 
of a sine wave with amplitude $\varepsilon$ and frequency $\nu$, i.e. `
\begin{equation}
\tau(t)=\tau+\varepsilon\sin(\nu t).
\end{equation} 
The corresponding calculations of the suppression coefficient 
in the plane spanned by the control parameters 
$K$ and $\tau/T$ are shown in Fig. \ref{sl39}.
The respective values for the amplitude and the frequency of the 
delay modulation are $\varepsilon=40$ and $\nu=10$.
We notice a substantial enlargement of the suppression domain 
in comparison to the one for the constant time delay shown in 
Fig. \ref{sl38}. At the same time one observes larger values for the suppression coefficient, with the maximum in the depicted range of control parameters reaching values close to 20, but also an undesirable effect shows up in the form of increase of the minimal feedback gain $K$ necessary to obtain higher suppression factor.

\section{Conclusion}

It is thought that specific functional states of neural networks are
characterized by modulation of oscillatory activity in specific frequency
bands through basal ganglia-cortical loops. It is being confirmed by
recordings of the local field potential that the globus pallidus in
mammalian brains plays the role  of pacemaker, namely its cells are
constantly firing at precise frequency and through their cortical loops
these cells participate in a very important control and have a 
hierarchical frequency organization of the local field potential which can be related 
to the mean field used in our model.

The present study of the behaviour of large number of interacting oscillators coupled through their mean field shows that their synchronization can be diminished by using time-delay feedback of Pyragas type in much larger domain in the parameter space if the delay-time changes as the time goes on. This suggests that it is worthwhile to look for an optimal choice in the delay modulation in order to increase the suppression coefficient and to extend toward lower gains the domain with successful desynchronization. Further generalizations could be of interest such as introduction of several feedback terms with independent delays or feedback terms with multiple delays \cite{SOC08,AP08}. Another extension would be study of systems of non-identical oscillators which would be more appropriate model for real systems. An analytical description of the domain of suppression in this type of coupled oscillators would be of also interest for future studies.

\end{document}